\documentclass[preprint,superscriptaddress,aps,prl]{revtex4}

\usepackage{graphicx}
\usepackage{dcolumn}
\usepackage{bm}

\begin{document}


\title{Charge Distributions in Metallic Alloys:\\ a Charge Excess Functional 
theory approach}

\author{Ezio Bruno}
\email{ebruno@unime.it}
\affiliation{ Dipartimento di Fisica and Unit{\`{a}} INFM,
Universit{\`{a}} di Messina, Salita Sperone 31, 98166 Messina, Italy}
\author{Leon Zingales}
\affiliation{ Dipartimento di Fisica and Unit{\`{a}} INFM,
Universit{\`{a}} di Messina, Salita Sperone 31, 98166 Messina, Italy}
\author{Yang Wang}
\affiliation{Pittsburgh Supercomputing Center, Pittsburgh, PA 15213 USA}

\date{\today}

\begin{abstract}
The distribution of local charge excesses (DLC) in metallic alloys, previously
obtained as a result of the analysis of order N electronic structure 
calculations, is derived from a variational principle. 
A phenomenological Charge Excess Functional (CEF) theory is obtained which 
is determined by three concentration dependent, material specific, parameters that 
can be obtained from {\it ab initio} calculations.
The theory requires modest computational efforts and
reproduces with an excellent accuracy the DLC and the electrostatic energies 
of ordered, substitutionally disordered or segregating metallic alloys and,
hence, can be considered an efficient approach alternative to conventional
electronic structure calculations. The substantial 
reduction of computing time opens new perspectives for the understanding of 
metallic systems and their mechanical properties. 
\end{abstract}
\maketitle

The application of metallic alloys in a huge variety of high-tech areas,
ranging from medical prosthesis~\cite{medicals} to jet engines~\cite{jets}, 
requires a careful assessment of the mechanical properties of these materials. 
The determination of their phase diagrams is crucial to this end since 
the performances of alloys are heavily 
influenced by the various {\it crystalline phases} of which they are 
made~\cite{Cahn,tballoys,Pettifor,Kraft}.
Recently, Zhao~\cite{Zhao} introduced an innovative experimental
method that allows for the {\it rapid} yet {\it accurate} assessment of alloy 
phase diagrams. However, theoretical methods 
with similar {\it high-throughput} performances seem much beyond the status of art. 

Current theoretical approaches can be divided in two different classes. 
Theories in the first group are based on Landau's idea that ordering
occurs in alloys due to some instability of the high-T
solid solution phase (HTSSP) with respect to certain concentration 
fluctuations~\cite{Khachaturyan,s2}.
Such schemes often use the Coherent Potential Approximation (CPA) solution
for the HTSSP as the reference state for a perturbation theory. They can be
put in a very elegant form using a reciprocal space formalism, making
contact with the concept of long-range order (LRO) and with the 
Fermi surfaces properties~\cite{physrep,cupdacar}. In spite of many successes, 
these approaches suffered from the criticisms moved to
the CPA which, in its standard implementation within the density functional
theory (DFT), fails to account for the electrostatic energies in 
the HTSSP. Theories in the second class are based on Ising hamiltonians.
The corresponding parameters can be extracted from the experiment, 
from a perturbative expansion of the CPA HTSSP solution~\cite{GPM} or, as it is 
nowadays more common, from the total energies of various alloy 
{\it configurations} as obtained by DFT calculations~\cite{Colinet}. 
In this case, accuracy is controlled by the cut-off length $l$ assumed for 
Ising interactions. In metallic alloys LRO and 
short-range order (SRO) are entangled in such a way that the convergence 
with $l$ is slow, while DFT calculations would be required for about $N!$ 
alloy configurations~\cite{Franceschetti,Sanati}, with $N \approx l^3$. In turn, DFT 
calculations {\it for a single configuration} constitute a bottleneck: using
the fastest available algorithms~\cite{LSMS,LSGF} the number of 
floating-point operations required is proportional to $N$, 
with huge prefactors~\cite{note1}.
Remarkable results have been obtained by a mixed approach~\cite{Zunger} that 
includes the summation of series both in the direct and the reciprocal
spaces. However, in this approach no estimate of the size of the truncation 
errors is available.

In the present Letter we shall show that the computational labour necessary 
to obtain total energies for {\it any} alloy configuration can be greatly 
reduced. At variance of existing simplified approaches~\cite{tballoys,Franceschetti},  
we shall maintain accuracies comparable to those of DFT calculations. 
   
The  analysis~\cite{FWS,Abrikosov_cpa,Ruban1} of DFT
calculations data in extended metallic systems indicates 
that the net charges at the crystal sites, $q_i$, and the site Madelung 
potentials, $V_i$, are strongly correlated. More precisely~\cite{FWS}, 
{\it within numerical errors}, 
i) the DLC is {\it continuous} over a certain interval and ii) the pairs 
$(q_i,V_i)$, lie on {\it straight lines}, one for each alloying 
species. For future reference, the statement (ii) is conveniently rewritten 
as
\begin{equation}
\label{qvsv}
a_i q_i + V_i = k_i  
\end{equation}
For a specified configuration of the binary alloy A$_{c_A}$B$_{c_B}$, the 
coefficients in Eq.~(\ref{qvsv}) take the values $a_A$ 
and $k_A$ if the i-th site is occupied by an A atom or $a_B$ and $k_B$ 
otherwise. The results (i) and (ii), in the following referred to as the $qV$ laws,
have been numerically obtained within the Local Density Approximation (LDA) and 
the muffin-tin or the atomic sphere approximations
for the crystal potential~\cite{Ruban1,CPALF}
but not yet formally derived. Here their validity  
shall be assumed as an "empirical" evidence. 

Accurate calculations of the alloy total energies must necessarily 
keep into account the above results. It has been shown recently that an 
isomorphous CPA model including local fields (CPA+LF)~\cite{CPALF} and a 
modified screened CPA approach~\cite{Ruban1} are able to catch the linear nature 
of the $qV$ laws and compatible with (i). However, the same models are not 
able to self-consistently determine the DLC. Here we shall obtain the DLC
from a variational principle in terms of {\it one} electronic 
degree of freedom for each atom, the local charge excess $q_i$. In the 
resulting phenomenological Ginzburg-Landau theory, hereafter referred to as 
CEF, the site chemical occupations determine the $q_i$ at each site and the 
electrostatic energy. For all the (ordered or disordered) alloy configurations 
corresponding to the same mean concentration the theory is 
completely determined by {\it three} material specific parameters that can be 
obtained {\it ab initio} by supercell DFT calculations or by the CPA+LF theory.
In the following the CEF theory shall be presented and tested vs. order 
N Locally Self-consistent Multiple Scattering (LSMS) DFT calculations~\cite{PCPA2001} 
for bcc CuZn alloys. 

As in the case of LSMS calculations~\cite{LSMS}, we study the binary 
alloy A$_{c_A}$B$_{c_B}$, $c_A+c_B=1$, by 
the means of supercells of volume $V$ containing N atoms with periodic 
boundary conditions. Each site can be occupied by an A or a B 
atom. The nuclear charge $Z_i$ and 
the volume~\footnote{
There is some arbitrariety in the way in which the $\omega_i$ 
can be chosen: they could be built using the Wigner-Seitz 
construction (and possibly approximated by spheres), or they could be 
non-overlapping muffin-tin spheres to which an appropriate fraction 
of the interstitial volume is added.} 
$\omega_i$, $\sum_i \omega_i=V$, are associated 
with each crystal site. Each alloy configuration is specified by a set of 
occupation numbers~\footnote{
In our notation Latin indices identify the sites in the supercell and Greek 
indices the chemical species. In the sums, Latin indices 
run from $1$ to N and Greek indices take the values A and B.
}, $X_i^\alpha$, where $X_i^\alpha=1$ if the 
$i$-th site is occupied by a $\alpha$ atom or $0$ otherwise and 
$\sum_i X_i^\alpha=N c_\alpha$. The site 
charge excesses, $q_i=\int_{\omega_i} d\vec{r} \rho(\vec{r}) - Z_i$,
defined in terms of the electronic density $\rho(\vec{r})$, satisfy
the global electroneutrality constraint
\begin{equation}
\label{electroneutrality}
\sum_i q_i =0
\end{equation}
As it is verified within spherical approximations for the crystal potential and the 
LDA, we assume that the system total electronic energy consists of the sum of 
site-diagonal terms plus a Madelung term~\cite{Janak,FWS},
$E_M=\sum_{ij} M_{ij} q_i q_j=\frac{1}{2}\sum_i q_i V_i$.
The Madelung matrix elements $M_{ij}$ are defined as usual~\cite{Ziman,CPALF}, and
the Madelung potentials are given by,
\begin{equation}
\label{vi}
V_i=2 \sum_j M_{ij} q_j 
\end{equation}
Unless otherwise stated we use atomic units in which $e^2=2$.

We wish to develop a theory that determines the $q_i$ and incorporates the $qV$ 
laws. For a particular alloy configuration, Eq.~(\ref{electroneutrality}) and the 
substitution of Eqs.~(\ref{vi}) in Eqs.~(\ref{qvsv}) give $N+1$ linear equations 
in the $q_i$. This implies that not all the coefficients $a_\alpha$ and $k_\alpha$ 
can be independent. The CPA+LF model gives a hint for the missing relationship. 
In this theory~\cite{CPALF} the quantities $1/a_\alpha$ depend {\it only} on the mean 
alloy concentration and are viewed as the {\it responses} of impurity sites, embedded 
in the CPA 'mean' alloy, to a local field designed to simulate the Madelung 
potential. Furthermore, the CPA 'electroneutrality' condition 
for the zero-field charges, $b_\alpha=k_\alpha/a_\alpha$ gives  $c_A b_A+c_B b_B=0$. 
These circumstances 
suggest that the theory we are elaborating should allow for a possible 
renormalization of the constants $k_\alpha$ in different configurations 
corresponding to the same mean alloy concentration.

To make further progresses, we consider the Ginzburg-Landau functional 
of the site charge excesses, 
\begin{equation}
\label{functional}
\Omega([q],\mu)=\sum_i \frac{a_i}{2} (q_i-b_i)^2 + \sum_{ij} M_{ij} 
q_i q_j - \mu \sum_i q_i
\end{equation}
where the Lagrange multiplier $\mu$ has been introduced to impose global 
electroneutrality. The minimization of 
Eq.~(\ref{functional}) with respect to the order parameter field 
$\{q_i\}$ and to $\mu$ gives the set of Euler-Lagrange 
equations constituted by Eq.~(\ref{electroneutrality}) and by
\begin{equation}
\label{el1}
a_i (q_i-b_i) + 2 \sum_j M_{ij} q_j= \mu 
\end{equation}
Eqs.~(\ref{el1}) are equivalent to Eqs.~(\ref{qvsv}) only when $\mu=0$. 
When $\mu \ne 0$, the renormalization  
$k_\alpha \rightarrow k_\alpha +\mu$ occurs to ensure global 
electroneutrality. Once the four 
constants $a_\alpha$, $b_\alpha$ have been obtained for {\it a 
given alloy configuration}, they also can be used for other configurations. 
As we shall see, the {\it transferability} of the parameters within fixed concentration 
ensembles is a peculiar strength of the present approach.

Now, since $\Omega$ has the dimension of an energy and contains the 
electrostatic contribution $E_M$, we assume that, except 
but for an additive constant, its minimum value corresponds to the 
total electronic energy of the alloy configuration at hand. The 
quadratic terms in Eq.~(\ref{functional}) can be interpreted as 
energetic contributions due to local charge rearrangements and
$\mu$ as the chemical potential ruling  charge transfers. 

The explicit solution of the problem can be written in terms of
{\boldmath $\Lambda$}, the inverse of the matrix of elements
$2 M_{ij}+\sum_\alpha a_\alpha X_i^\alpha \; \delta_{ij}$. 
Straightforward calculations show that 
\begin{equation}
\label{solqfinal}
q_i= (k_A - k_B) \; [ (1-y) \sum_j
\Lambda_{ij} {X_j}^A - y  \sum_j \Lambda_{ij}{X_j}^B ]
\end{equation}
where $\Lambda_{\alpha\beta} =\sum_{ij} {X_i}^\alpha \Lambda_{ij} 
{X_j}^\beta $, and 
$y = \sum_\alpha \Lambda_{A\alpha}/\sum_{\alpha\beta}\Lambda_{\alpha\beta}$.
The set of four 
constants found by LSMS or similar calculations for a specific configuration
is then equivalent to the three constants in the CEF theory, $a_A$, $a_B$ and 
$k_A-k_B$, and to the chemical potential $\mu$.
The CEF formulation and solution are well defined for both ordered 
and disordered alloys.

\begin{table}
\caption{Characterization of the samples used and CEF parameters, 
a$_{Cu}$, a$_{Zn}$ and k$_{Cu}$-k$_{Zn}$. All supercells correspond to
bcc equiatomic CuZn alloys at the lattice constant a=5.5 a.u.
SRO are the Warren-Cowley short-range order parameters for the first 3 
neighbours shells. 
Samples 1-3 and samples 4-5 correspond, respectively, 
to random and partially ordered configurations, samples 6 and 7 are, 
respectively, Cu/Zn and  Cu$_{0.125}$Zn$_{0.875}$/Cu$_{0.875}$Zn$_{0.125}$ 
multilayers stacked along (001). LSMS calculations for sample~1 have 
been published in Ref.~\onlinecite{PCPA2001}. 
}
\begin{ruledtabular}
\begin{tabular}{cccccrrr}
Samples & N & a$_{Cu}$ & a$_{Zn}$ & k$_{Cu}$-k$_{Zn}$ & & SRO & \\
\hline
 1 & 1024 & 1.84 & 1.82 & 0.29 & -0.004 &  0.021 &  0.010 \\
 2 &  256 & 1.85 & 1.83 & 0.29 &  0.078 &  0.042 &  0.021   \\
 3 &  256 & 1.81 & 1.80 & 0.28 &  0.031 & -0.005 & -0.021  \\
 4 &  256 & 1.83 & 1.83 & 0.29 &  0.184 &  0.120 & -0.078  \\
 5 &  256 & 1.90 & 1.87 & 0.30 & -0.309 &  0.271 &  0.141  \\
 6 &  256 & 1.81 & 1.83 & 0.28 &  0.750 &  0.833 & 0.667   \\
 7 &  256 & 1.85 & 1.85 & 0.29 &  0.414 &  0.458 & 0.375 \\
\end{tabular}
\label{tabI}
\end{ruledtabular}
\end{table}
In the following we shall apply the CEF theory  
to bcc Cu$_{0.50}$Zn$_{0.50}$ alloys and compare 
the results vs. LSMS calculations. We have selected several sample
supercells described in Table I and designed to simulate random, 
partially ordered or segregated configurations. 
For each sample a set of the CEF parameters has been 
extracted by linear fits of the corresponding LSMS $qV$ data: CEF 
calculations made using parameters from the $n$-th sample are indicated 
below as CEF-$n$. 

\begin{table}
\caption{CEF-1 calculations for the 
bcc Cu$_{0.50}$Zn$_{0.50}$ random alloy sample 1 compared with 
the LSMS results of Ref.~\onlinecite{PCPA2001}. $\langle q \rangle_{Cu}$ and  $\langle 
V \rangle_{Cu}$ are, respectively, the mean charges and  
Madelung potentials at the Cu sites, $\sigma_{Cu}$ and $\sigma_{Zn}$ the 
standard deviations of the DLC's for Cu and Zn sites, $E_M/N$ 
is the Madelung energy per atom and $\langle (\Delta q)^2 \rangle$ are the 
mean square deviations between CEF and LSMS charges.}
\begin{ruledtabular}
\begin{tabular}{ccc}
                            & CEF-1           & LSMS  \\   
\hline
$\langle q \rangle_{Cu}$    & 0.099787   & 0.099783   \\
$\langle V \rangle_{Cu}$ & -0.038197     & -0.038188   \\
$\sigma_{Cu}$               & 0.02507    & 0.02523    \\
$\sigma_{Zn}$               & 0.02801    & 0.02814    \\
$E_M/N$   (mRy)         & -2.552         & -2.557     \\
$\langle (\Delta q)^2 \rangle$ & 2.7 10$^{-6}$ &             \\
\end{tabular}
\end{ruledtabular}
\label{tabII}
\end{table}

\begin{figure}
\includegraphics[width=7cm]{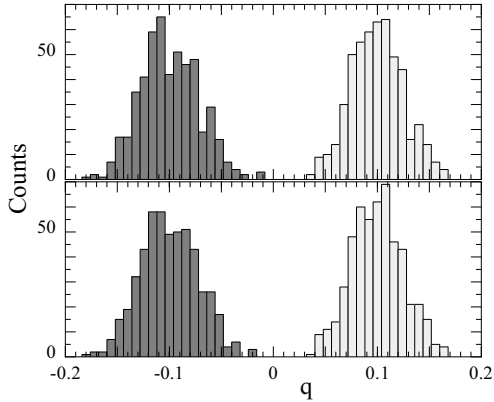} 
\caption{Cu (light histogram) and Zn (dark histogram) calculated DLC for the bcc 
Cu$_{0.50}$Zn$_{0.50}$ random alloy sample 1. 
Top frame: LSMS (Ref.~\onlinecite{PCPA2001}); lower frame: CEF-1.}
\label{histo}
\end{figure}

A detailed comparison between LSMS and CEF-1 calculations for sample 1 is 
reported in Table II. The two sets of calculations present very small differences:
5 parts over $10^5$ for the mean values of the charges and of the 
Madelung potentials, 2 parts over $10^4$ for the Madelung energies, less 
than 1 per cent for the widths of the DLC.  The absolute values of the 
differences 
$\Delta q_i=q_i^{CEF-1}-q_i^{LSMS}$ are smaller than 0.005 electrons 
at any lattice site and not correlated with the chemical occupations. 
The mean square deviation between 
the two set of charges, $\langle (\Delta q)^2 \rangle$, is of the order of 
$10^{-6}$, i.e. it is comparable with the numerical errors in LSMS 
calculations. The resulting DLC's, plotted in Fig.~(\ref{histo}), appear very 
similar.
The main source of the tiny differences is that all the CEF charges
{\it by construction} satisfy the $qV$ laws, while the same laws hold 
only approximately for LSMS calculations. 

Specific tests have been designed
to check the {\it transferability} of the CEF parameters extracted 
from one sample to the others.
The mean square deviations of the charges obtained from CEF-$n$ 
and LSMS $\langle (\Delta q)^2 \rangle$ (Table III) clearly show that 
there is no appreciable loss of accuracy 
when the CEF parameters extracted from random samples are used in ordered,
partially ordered or segregated samples and vice versa.  
Calculations~\cite{BZWunpub} for CuPd alloys support the same conclusion. The 
{\it transferability} of the CEF parameters is a very remarkable result and 
implies that the theory is generally applicable to 
metallic alloys, no matters whether they are ordered, 
disordered or segregated. Furthermore, the very high accuracy obtained for 
the Madelung energies (Table IV) indicates that the theory can 
describe very carefully the electrostatic contributions to the energetics of 
ordering phenomena.

\begin{table}
\caption{Charge mean square deviations $\langle(\Delta q)^2 \rangle 
\times 10^6$ 
between CEF-$n$ and LSMS calculations. Columns identify different samples.}
\begin{ruledtabular}
\begin{tabular}{crrrrrrr}
         &1    & 2   & 3   & 4   & 5 & 6   & 7 \\
\hline
 CEF-1   & 3   & 3   & 3   & 2   & 7 & 0.4 & 3 \\
 CEF-2   & 3   & 3   & 3   & 2   & 7 & 0.4 & 3 \\
 CEF-3   & 3   & 2   & 3   & 1   & 9 & 0.5 & 3 \\
 CEF-4   & 3   & 3   & 3   & 2   & 7 & 0.4 & 2 \\
 CEF-5   & 4   & 4   & 5   & 3   & 4 & 0.5 & 3 \\
 CEF-6   & 2   & 3   & 3   & 2   & 6 & 0.5 & 2 \\
 CEF-7   & 4   & 4   & 4   & 2   & 5 & 0.4 & 3 \\
\end{tabular}
\end{ruledtabular}
\label{tabIII}
\end{table}

\begin{table}
\caption{Madelung energies per atom (in mRyd units) from LSMS and CEF calculations. 
Columns identify different samples.}
\begin{ruledtabular}
\begin{tabular}{cccccccc}
 &1 & 2  & 3     & 4       & 5        & 6  & 7\\
\hline
 LSMS   & -2.56 & -2.15 & -2.32 & -1.42 & -3.70  &  0.20  & -1.02 \\
 CEF-1  & -2.55 & -2.11 & -2.30 & -1.40 & -3.82  &  0.21  & -1.03 \\
 CEF-2  & -2.54 & -2.10 & -2.29 & -1.40 & -3.80  &  0.22  & -1.02 \\
 CEF-3  & -2.57 & -2.12 & -2.31 & -1.41 & -3.85  &  0.20  & -1.05 \\
 CEF-4  & -2.54 & -2.10 & -2.29 & -1.40 & -3.81  &  0.21  & -1.03 \\
 CEF-5  & -2.49 & -2.05 & -2.24 & -1.37 & -3.71  &  0.25  & -0.98 \\
 CEF-6  & -2.52 & -2.08 & -2.27 & -1.39 & -3.78  &  0.20  & -1.02 \\
 CEF-7  & -2.50 & -2.06 & -2.25 & -1.37 & -3.73  &  0.23  & -1.00 \\
\end{tabular}
\end{ruledtabular}
\label{tabIV}
\end{table}

We conclude this Letter with remarks about interesting aspects of the CEF 
model in view of possible future applications.

The CEF operates a {\it coarse graining} over the electronic degrees 
of freedom that are reduced to {\it one for each atom}, the local excess of 
charge. This notwithstanding, the theory carefully reproduces the 
DLC and the energetics of metallic alloys in any ordering status. 
This has been possible because the Madelung potentials, through 
Eq.~(\ref{vi}), weight as appropriate the {\it long ranged} effects of the 
occupations of {\it all} the crystal sites. Such a renormalization of the 
interactions holds {\it exactly} within CPA-based theories like the 
CPA+LF~\cite{CPALF} or the PCPA~\cite{Ujfalussy}, where any site diagonal 
property is a {\it unique function} of the Madelung potential $V_i$ and the 
nuclear charge $Z_i$ at the same site. This uniqueness does not 
hold~\cite{CPALF} for more exact approaches, where some residual dependence 
on the site nearest neighbours environment is expected for. Nevertheless, 
the fact that CPA-based theories accurately accounts for the spectral 
properties of metallic alloys~\cite{Abrikosov_cpa} and the quantitative 
agreement with LSMS calculations, in the present work as well as in 
Refs.~\onlinecite{CPALF} and~\onlinecite{Ujfalussy}, 
suggest that the errors introduced by neglecting the nearest neighbours 
effects not already conveyed by the $V_i$ are comparable with numerical 
errors in DFT calculations. We mention that, in a very different 
context, precedents of this idea of a coarse graining over quantum 
degrees of freedom can be found in the concepts of chemical valence and 
of electronegativity.  

The calculations presented in this Letter require $N^3$ floating-point
operations in order to obtain  
{\boldmath $\Lambda$} by conventional linear algebra algorithms. Thus, 
for $N=1000$, the CEF is about $10^4$ times faster than LSMS~\cite{note1}. We are 
convinced that, in the next future, such a large computational speed 
up and the mentioned transferability of the CEF parameters will make 
possible the development of accurate ab initio techniques for
the investigation of ordering phenomena and the calculation of phase 
diagrams in metallic alloys.

We thank J.S. Faulkner who made the data of Ref.~\onlinecite{PCPA2001} available 
in digital form  and acknowledge discussions with E.S. Giuliano.


\begin{thebibliography}{24}
\bibitem{note1}
In Ref.~\onlinecite{LSMS} the 
size of these prefactors is estimated of the order of $10^{10}$.
\end{thebibliography}


\end{document}